\documentclass[traditabstract]{aa}

\usepackage{graphicx,amssymb,amsmath}
\usepackage{txfonts}
\usepackage{natbib}

\newcommand{\argsh}{{\rm \, arcsinh \,}}
\newcommand{\elie}{{\mathbf E}}

\newcommand{\bara}{\bar{a}}
\newcommand{\rint}{R_{\rm in}}
\newcommand{\rext}{R_{\rm out}}
\newcommand{\ag}{\langle a \rangle}
\newcommand{\aout}{a_2}
\newcommand{\ain}{a_1}
\newcommand{\elik}{{\mathbf K}}

\newcommand{\so}{\Sigma_{0}}
\newcommand{\psic}{\psi_{\rm c}}
\newcommand{\psid}{\psi_{\rm d}}
\newcommand{\psihs}{\psi_{\rm hs}}

\newcommand{\rhom}{\rho_{\rm magic}}

\newcommand{\rmin}{R_{\rm min}}
\newcommand{\mmin}{m_1}
\newcommand{\mmout}{m_2}
\newcommand{\mpin}{m'_1}
\newcommand{\mpout}{m'_2}

\newcommand{\din}{\Delta_1}
\newcommand{\dout}{\Delta_2}

\begin{document}

\title{Self-gravity at the scale of the polar cell}

\titlerunning{Self-gravity at the scale of the polar cell}

\author{Jean-Marc Hur\'e\inst{1,2}, Arnaud Pierens\inst{3} \and Franck Hersant\inst{1,2} }

\offprints{jean-marc.hure@obs.u-bordeaux1.fr}

\institute{Universit\'e de Bordeaux, Observatoire Aquitain des Sciences de l'Univers
\and
CNRS/INSU/UMR 5804/LAB, BP 89, 33271 Floirac cedex, France
\and
LAL-IMCCE/USTL, 1 Impasse de l'Observatoire, F-59000 Lille, France}


\date{Received 09/02/2009 / Accepted 05/03/2009}

\abstract
{We present the exact calculus of the gravitational potential and acceleration along the symmetry axis of a plane, homogeneous, polar cell as a function of mean radius $\bara$, radial extension $\Delta a$, and opening angle $\Delta \phi$. Accurate approximations are derived in the limit of high numerical resolution at the geometrical mean $\ag$ of the inner and outer radii (a key-position in current FFT-based Poisson solvers). Our results are the full extension of the approximate formula given in the textbook of Binney \& Tremaine to all resolutions. We also clarify definitely the question about the existence (or not) of self-forces in polar cells. We find that there is always a self-force at radius $\ag$ {\it except if the shape factor} $\rho \equiv \bara \Delta \phi /\Delta a \rightarrow 3.531$, asymptotically. Such cells are therefore well suited to build a polar mesh for high resolution simulations of self-gravitating media in two dimensions. A by-product of this study is a newly discovered indefinite integral involving complete elliptic integral of the first kind over modulus.}

\maketitle

\section{Introduction}
\label{sec:intro}

The structure and dynamical evolution of astrophysical discs is mainly governed by gravity. In this context, the numerical computation of the gravitational potential and forces of discs is of fundamental importance and the improvement in accuracy, resolution and computing time remains an interesting challenge \citep[e.g.,][]{must95,maha03,londrillo04,hure05,jusu07,lietal08}. In many (if not all) hydrodynamical simulations of flat, self-gravitating discs \citep[e.g.,][]{hupf01,zhang08,bm08}, the potential is derived everywhere in the disc by means of Fast Fourier Transforms \citep[e.g.,][]{binneytremaine87}. This is an efficient technique, which can be extended to deduce directly accelerations \citep{bm08}. A weak point however is in the determination of the self-gravitating component (i.e., the effect of a cell on itself), which involves an improper integral. An approximation of this self-potential was proposed in \cite{binneytremaine87}, but for a cell in which the surface density varies with the polar radius $R$ as $R^{-3/2}$. In the same conditions, \cite{bm08} argued that the radial acceleration is expected to vanish at the center of each computational cell. This assertion is certainly correct asymptotically for homogeneous cells as their size becomes smaller and smaller.
We stress that because Newton's law goes like $R^{-2}$, the self-gravitating component generally provides a significant contribution to the total potential or force, and must therefore be treated as properly as possible, especially at high resolution. Any error, even relatively small, may introduce artifacts or biases in models.  

In this letter, we report the exact expressions for both the potential and radial acceleration due to a homogeneous polar cell inside the cell itself and study the high-resolution limit. We recall in Sect. \ref{sec:polarmesh} the integral expression for the potential along the axis of a polar cell which can be decomposed into two parts: a potential created by an homogeneous disc and the other a potential due to horseshoe-like surface. Sections \ref{sec:disc} and \ref{sec:hs} are devoted to the determinatons of these two contibutions. The results for the polar cell (potential and acceleration) are obtained in Sect. \ref{sec:res}. We discuss the case of high resolutions in Sect. \ref{sec:highres} and show that there is always a self-acceleration except if the cell has a special shape. These results are important since they contribute to the reliability and accuracy of current simulations of self-gravitating media on a polar mesh.
 
\section{The polar mesh}
\label{sec:polarmesh}

\begin{figure}
\includegraphics[width=8.9cm]{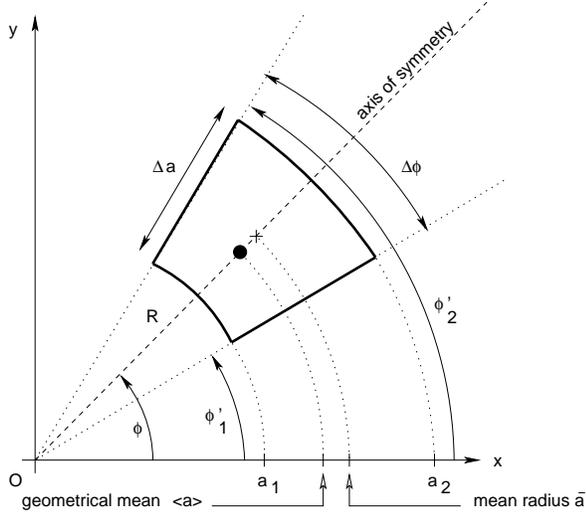}
\caption{A polar cell with inner radius $\ain$, outer radius $\aout$, radial extension $\Delta a = \aout - \ain$, and opening angle $\Delta \phi$.}
\label{fig:cell.eps}
\end{figure}

\begin{figure}[h]
\includegraphics[width=8.9cm]{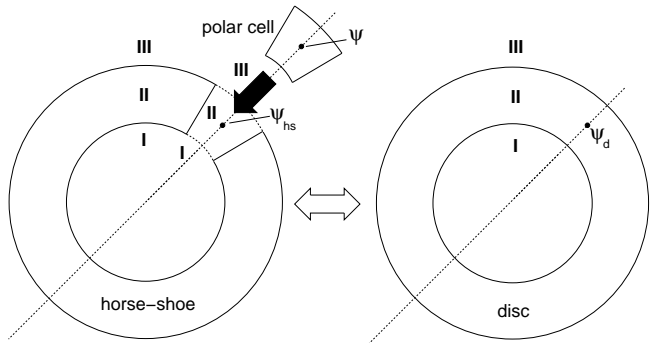}
\caption{A polar cell is the superposition of a disc and a horseshoe with negative surface density.}
\label{fig:hshoe.eps}
\end{figure}

We consider a planar, homogeneous, polar cell with inner polar radius $\ain$, outer radius $\aout=\ain+\Delta a$, lower azimuth $\phi'_1$, and upper azimuth $\phi'_2=\phi'_1+ \Delta \phi$ as shown in Fig. \ref{fig:cell.eps}. At any position $(R,\phi)$ in the plane of this cell, the gravitational potential is given by Newton's generalized formula:
\begin{equation}
\psi(R,\phi)=- G \so \int_{\ain}^{\aout}{\int_{\phi'_1}^{\phi'_2}{\frac{a da d\phi' }{\sqrt{a^2+R^2-2aR\cos(\phi'-\phi)} }}},
\end{equation}
where $\so$ is the surface density. This cell has a single axis of symmetry defined by $\phi=\frac{1}{2}(\phi'_1+\phi'_2)$. If we introduce the quantity:
\begin{equation}
m=\frac{2\sqrt{aR}}{a+R}
\label{eq:mmodulus}
\end{equation}
as well as the new variable $\theta'$ such that $2\theta'=\pi - (\phi'-\phi)$, the potential {\it along its axis of symmetry} is given by:
\begin{equation}
\psi(R)=- 2G \so \int_{\ain}^{\aout}{{\int_{\theta}^{\frac{\pi}{2}}{\sqrt{\frac{a}{R}} m \frac{ da d\theta'}{\sqrt{1-m^2 \sin^2 \theta'}}}}},
\end{equation}
where $4\theta = 2 \pi-\Delta \phi$. The integral over $\theta'$ can be decomposed into two integrals with $0$ as lower bound. We then have:
\begin{equation}
\psi(R)=- 2G \so \int_{\ain}^{\aout}{da\sqrt{\frac{a}{R}}m{\left[\elik(m) - F(\theta,m)\right]}},
\label{eq:psialongaxis}
\end{equation}
where
\begin{equation}
F(\theta,m) = \int_0^\theta{\frac{d\theta'}{\sqrt{1-m^2 \sin^2 \theta'}}}
\end{equation}
is the incomplete elliptic integral of the first kind, $m$ is the modulus, $\theta$ is the amplitude, and $\elik(m)~\equiv~F(\frac{\pi}{2},m)$ is the complete elliptic integral of the first kind. The interpretation of this relation is presented in Fig. \ref{fig:hshoe.eps} and follows from the superposition principle: the first part of the integral of Eq. \ref{eq:psialongaxis} is the potential due to a flat disc, namely:
\begin{equation}
\psid(R)= - 2G \so \int_{\ain}^{\aout}{da\sqrt{\frac{a}{R}}m\elik(m)},
\label{eq:superI}
\end{equation}
 and the second part is that of a flat, horseshoe-like surface:
\begin{equation}
\psihs(R) = - 2G \so \int_{\ain}^{\aout}{da\sqrt{\frac{a}{R}}mF(\theta,m)}.
\label{eq:superII}
\end{equation}

We can easily derived $a$ as a function of $m$ from Eq. \ref{eq:mmodulus}, as well as the corresponding derivative $da/dm$. We have, respectively:
\begin{equation}
\frac{a}{R} = \left( \frac{1 \pm m'}{m}\right)^2
\label{eq:avsm}
\end{equation}
and
\begin{equation}
\pm m' \frac{da}{2R} = - \frac{(1 \pm m' )^2}{m^3} dm,
\label{eq:da}
\end{equation}
where $m'~=~\sqrt{1-m^2}$ is the complementary modulus, $+m'$ is for $R \le a$ (i.e., matter is located right to the position $R$) and $-m'$ is for $R \ge a$ (i.e., matter is located to the left). From Eqs.(\ref{eq:mmodulus}) and (\ref{eq:da}), we deduce that the self-gravitating potential in the cell along its symmetry axis is:
\begin{equation}
\psi(R)= 4G \so R \int_{\mmin}^{\mmout}{\frac{(1 \pm m' )^3}{\pm m' m^3} \left[\elik(m) - F(\theta,m)\right] dm}.
\label{eq:psidm}
\end{equation}

We note that this expression is rarely seen in the literature.

\section{The homogeneous disc}
\label{sec:disc}

\subsection{Classical derivation}

To calculate $\psid$, one classically changes\footnote{The transformation is \citep[e.g.][]{gradryz65}:
\begin{equation}
K\left(\frac{2\sqrt{x}}{1+x}\right)=(1+x)K\left(x\right).
\label{eq:changem}
\end{equation}} the modulus $m$ of $\elik$ by setting $m=\frac{2\sqrt{x}}{1+x}$ where $x \le 1$. The new modulus $x$ is either $a/R \equiv u \le 1$ or $R/a \equiv v \le 1$ depending on $R$. 
So, the potential due to an homogeneous disc is given by:
\begin{equation}
\frac{\psid(R)}{-4 G \so R} =
\begin{cases}
- \int_{v_1}^{v_2}{\elik(v)\frac{dv}{v^2}}, \qquad \text{in region (I)},\\
\int_{u_1}^1{u\elik(u)du} - \int_1^{v_2}{\frac{\elik(v)}{v^2}dv},\\ \qquad \text{in region (II)},\\
\int_{u_1}^{u_2}{u\elik(u)du}, \qquad \text{in region (III)},
\end{cases}
\label{eq:discuv}
\end{equation}
where region (I) is for $R \le \ain$, region (II) is for $ \ain \le R \le \aout$, and region (III) is for $R \ge \aout$ (see Fig. \ref{fig:hshoe.eps}). Since the indefinite integrals in Eqs.(\ref{eq:discuv}) are known\footnote{In particular, we have for any $k \le 1$ \citep[][]{gradryz65}:
\begin{equation}
\int{\elik(k)\frac{dk}{k^2}} = - \frac{\elie(k)}{k} \quad \text{and} \quad \int{\elik(k)k dk} = \elie(k)- {k'}^2\elik(k),
\label{eq:intek}
\end{equation}
where  $\elie$ is the complete elliptic integral of the second kind and $k'~=~\sqrt{1-k^2}$ is the complementary modulus.}, a close-form expression for $\psid$ can finally be deduced. It is:
\begin{equation}
\frac{\psid(R)}{-4 G \so R} =
\begin{cases}
\left[\frac{\elie(v)}{v}\right]_{v_1}^{v_2} \quad \text{in (I)},\\
\left[\elie(u)-{u'}^2 \elik(u)\right]_{u_1}^1+\left[\frac{\elie(v)}{v}\right]_1^{v_2} \quad \text{in (II)},\\
\left[\elie(u)-{u'}^2 \elik(u)\right]_{u_1}^{u_2} \qquad \text{in (III)}.
\end{cases}
\label{eq:discuv_closed}
\end{equation}

\subsection{A new indefinite integral ?}

From a numerical point of view, it is preferable to use $u$ and $v$ as modulus of the complete elliptic integrals rather than $m$ (see  Sect. \ref{sec:highres}). However, an interesting issue is raised if we perform a change of modulus in Eqs.(\ref{eq:discuv_closed}), to restore $m$ as the modulus of $\elie$ and $\elik$. We find:
\begin{equation}
\label{eq:discm_closed}
\frac{\psid(R)}{-4 G \so R} =
\begin{cases}
\left[ \frac{\elie(m) + m' \elik(m)}{1 - m'} \right]_{\mmin}^{\mmout} \quad \text{in (I)},\\
\left[ \frac{\elie(m) -  m' \elik(m)}{1 + m'} \right]_{\mmin}^1 + \left[ \frac{\elie(m) + m' \elik(m)}{1 - m'} \right]_1^{\mmout} \, \text{in (II)},\\
\left[ \frac{\elie(m) -  m' \elik(m)}{1 + m'} \right]_{\mmin}^{\mmout} \qquad \text{in (III)}.
\end{cases}
\end{equation}
If we now compare these expressions with the first part of Eq. \ref{eq:psidm}, we conclude that:
\begin{equation}
\int{\frac{(1 \pm k' )^3}{k^3 k'} \elik(k) dk} = \mp \frac{\elie(k) \pm k' \elik(k)}{1 \mp k'},
\label{eq:newint}
\end{equation}
for any modulus $k$ and complementary modulus $k'$. 
 This indefinite integral is probably new \citep{jef09}.

\section{The homogeneous horseshoe}
\label{sec:hs}

To calculate  $\psihs$, we have attempted to proceed as for the disc, by applying a change of modulus. We failed, mainly because, as mentioned, the new amplitude $\alpha$ of $F$ resulting from Eq. \ref{eq:changem} now becomes a function of the modulus. This introduces a severe mathematical difficulty that we were unable to circumvent. The answer is however obtained by considering Eq. \ref{eq:newint} with incomplete elliptic integrals. Using the partial derivative\footnote{In particular, we have \citep[e.g.][]{gradryz65}:
\begin{equation}
\frac{\partial E}{\partial k} = \frac{E-F}{k} \quad \text{and} \quad \frac{\partial F}{\partial k} = \frac{E-{k'}^2 F}{k {k'}^2} - \frac{k \sin \theta \cos \theta}{{k'}^2 \sqrt{1- k^2 \sin^2 \theta}},
\label{eq:derivef}
\end{equation}
where $E(\theta,k)$ is the incomplete elliptic integral of the second kind.} of $F$ and $E$ with respect to their modulus (keeping the amplitude constant), we find that:
\begin{equation}
\frac{\partial }{\partial k} \left[\mp \frac{E(\theta,k) \pm k' F(\theta,k)}{1 \mp k'}\right] = \frac{(1 \pm k' )^3}{k^3 k'} F(\theta,k)  +  H(\theta,k)
\label{eq:hexists}
\end{equation}
where
\begin{equation}
H(\theta,k) = \frac{(1 \pm k')}{2 k k' \Delta} \sin 2 \theta
\label{eq:h}
\end{equation}
and  $\Delta = \sqrt{1-k^2 \sin^2 \theta}$. In fact, $\int{H(\theta,k) dk}$ can be expressed exactly in terms of basic functions.  Actually, after some algebra, we have:
\begin{equation}
\int{H(\theta,k)dk} = \frac{\sin 2\theta }{4} \left[\ln \frac{\Delta-k'}{\Delta+k'} \pm \ln \frac{1-\Delta}{1 + \Delta} \right].
\label{eq:inth}
\end{equation}
It follows from Eqs.(\ref{eq:psidm}), (\ref{eq:hexists}), (\ref{eq:h}), and (\ref{eq:inth}) that the potential along the axis of the homogeneous horseshoe is given by:
\begin{equation}
\frac{\psihs(R)}{- 4G \so R } =
\begin{cases}
 \left[ \frac{ E(\theta_2,m) + m' F(\theta,m)}{1 - m'} -  \frac{\sin 2\theta }{4} \ln \frac{(\Delta+m')(1 + \Delta)}{(\Delta-m')(1-\Delta)} \right]_{\mmin}^{\mmout},\\
 \qquad \text{in (I)},
\\
 \left[ \frac{ E(\theta,m) - m' F(\theta,m)}{1 + m'} + \frac{\sin 2 \theta }{4} \ln \frac{(\Delta+m')(1 - \Delta)}{(\Delta-m')(1+\Delta)} \right]_{\mmin}^1\\
\; + \left[ \frac{ E(\theta,m) + m' F(\theta,m)}{1 - m'} -  \frac{\sin 2\theta }{4} \ln \frac{(\Delta+m')(1 + \Delta)}{(\Delta-m')(1-\Delta)} \right]_1^{\mmout}\\
 \qquad \text{in (II)},
\\
 \left[ \frac{ E(\theta,m) - m' F(\theta,m)}{1 + m'} + \frac{\sin 2 \theta }{4} \ln \frac{(\Delta+m')(1 - \Delta)}{(\Delta-m')(1+\Delta)} \right]_{\mmin}^{\mmout},\\ \qquad \text{in (III)}.
\end{cases}
\label{eq:hsfinal}
\end{equation}
As for Eq. \ref{eq:discuv_closed}, the expression for region (II) can be simplified.

\begin{figure}
\includegraphics[width=8.9cm]{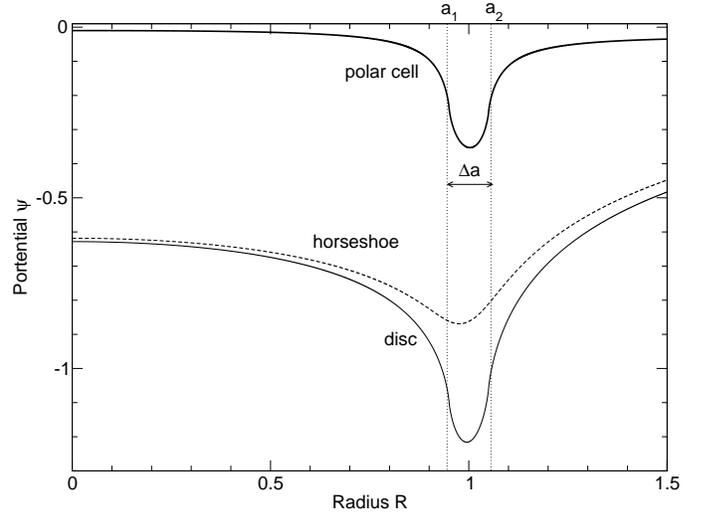}
\caption{The potential along the symmetry axis of a homogeneous polar cell with geometrical parameters $\ain=0.95$, $\Delta a=0.1$ and $\Delta \phi = 0.1$ ($\bara=1$ and $\ag\approx 0.99875$). The minimum lies at $\rmin \approx 0.9949$. Also shown are the potential of the associated disc and horseshoe.}
\label{fig:psi.eps}
\end{figure}

\section{Potential and acceleration in the polar cell}
\label{sec:res}

The potential inside a polar mesh and along its symmetry axis is found from Eqs.(\ref{eq:discm_closed}b) and (\ref{eq:hsfinal}b). It is:
\begin{flalign}
\nonumber
&\frac{\psi(R)}{- 4G \so R } = \frac{\elie(\mmout)-E(\theta,\mmout)+\mpout \left[\elik(\mmout)-F(\theta,\mmout)\right]}{1-\mpout}\\
\nonumber
& \qquad \qquad -  \frac{\elie(\mmin)-E(\theta,\mmin) - \mpin \left[\elik(\mmin)-F(\theta,\mmin)\right]}{1+\mpin}\\
& - \frac{\sin 2 \theta}{4} \ln \frac{(\din-\mpin)(1+\din)(\dout-\mpout)(1-\dout)}{(\din+\mpin)(1-\din)(\dout+\mpout)(1+\dout)},
\label{eq:pmfinalII}
\end{flalign}
where $\Delta_i^2 = 1 - m_i^2 \sin^2 \theta$ with $i=\{1,2\}$. This expression is exact. Figure \ref{fig:psi.eps} displays $\psid$, $\psihs$ and their difference $\psi$ versus $R$ for a typical cell. By deriving this formula with respect to $R$, we obtain the relation:
\begin{equation}
\frac{d\psi}{dR}= \frac{\psi}{R} + 2 G \so \left[  \frac{(1 \pm m' )^3}{m^2} \left\{ \elik(m)-F(\theta,m)\right\} \right]_{\mmin}^{\mmout}
\label{eq:odepmesh}
\end{equation}
which is the opposite of the gravitational acceleration $g_R$ along the cell axis. If we set $\varpi=R/\aout$ and $\tilde{\psi}=\psi/\psic$ where
\begin{equation}
\psic = - 2 \pi G \so \aout \left(1 - \frac{\ain}{\aout} \right) \left(1- \frac{2\theta}{\pi} \right)
\label{eq:psic}
\end{equation}
is the potential due to the cell at the origin $R=0$, and
\begin{equation}
S_\theta(\varpi)=  - \frac{\left[  \frac{(1 \pm m' )^3}{m^2} \left\{ \elik(m)-F(\theta,m)\right\} \right]_{\mmin}^{\mmout}}{\left(1-\frac{\ain}{\aout}\right)\left(\pi - 2\theta\right)},
\label{eq:unifieds}
\end{equation}
then Eq. \ref{eq:odepmesh} takes the form:
\begin{equation}
\frac{d \tilde{\psi}}{d \varpi}= \frac{\tilde{\psi}}{\varpi} + S_\theta(\varpi).
\label{eq:odepmesh_hh07like}
\end{equation}
This is the generalization of the Ordinary Differential Equation (ODE) derived by \cite{hh07} to a polar cell with opening angle $\Delta \phi = 2\pi - 4 \theta$ (which becomes a disc for $\theta~=~0$). This ODE can be easily solved numerically using standard schemes since $\psi$ is known both at the center (see Eq. \ref{eq:psic}) and at infinity where $\psi=0$. The full generalization to polar cells where the surface density $\Sigma$ is a power-law of the radius, i.e., $\Sigma(a) = \Sigma_0 (a/\aout)^s$, seems straightforward.

\section{Approximations at high numerical resolutions and the ``magic'' shape factor.}
\label{sec:highres}

At high resolution, polar cells tends to become segments, squares, or arcs depending on the shape factor $\rho$ of the cell defined by:
\begin{equation}
\bara \Delta \phi = \rho \Delta a,
\end{equation}
where $\bara=\frac{1}{2}(\ain+\aout)$ is the center of physical cells. When both $\Delta \phi \ll 1$ (i.e. $\theta \rightarrow \frac{\pi}{2}^-$) and $\Delta a \ll \bara$, we can expand $F(\theta,m)$, $E(\theta,m)$, $\elik(m)$, and $\elie(m)$ over $m'$ \citep[e.g.,][]{cg85,gradryz65} to obtain an approximation for $\psi$ and $g_R$ as a function of $R$ and the cell geometrical parameters $(\bara, \Delta a, \Delta \phi)$. 
A radius plays a key role in current FFT-based Poisson solvers (see references in Sect. \ref{sec:intro}): this is the geometrical mean of the inner and outer radii, namely $\ag=\sqrt{a_1 a_2}$. This radius is the center of computational cells when a radial logarithmic mapping is applied to the entire physical mesh. For $R = \ag$, the two modulus $m_1$ and $m_2$ are equal (and so are the complementary), and we have $m'_1 = m'_2 \approx \frac{\Delta a}{4 \bara}$ to first order.  After some algebra, we find:
\begin{equation}
\psi(\langle a\rangle) \approx - 2 G \so \Delta a \left[ \argsh \rho + \rho \argsh \frac{1}{\rho} \right],
\label{eq:approxpgeo}
\end{equation}
and
\begin{flalign}
\label{eq:approxgrgeo}
g_R(\ag) & \approx G \so \frac{\Delta a}{\bar{a}} \left( 2 \rho \argsh \frac{1}{\rho} -  \argsh \rho \right),
\end{flalign}

\begin{figure}[h]
\includegraphics[width=8.9cm]{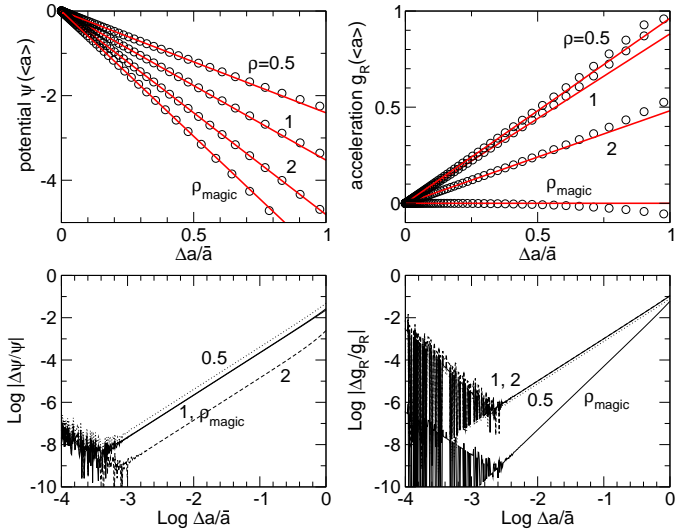}
\caption{{\it Top panels}: exact potential and acceleration in a polar cell at radius $R=\ag$ for three shape factors $\rho$ ({\it circles}) as well as their approximations ({\it lines}) by Eqs.(\ref{eq:approxpgeo}) and (\ref{eq:approxgrgeo}). {\it Bottom panels}: decimal logarithm of the relative error. Also displayed are the results obtained with the "magic" value $\rhom \approx 3.531$.}
\label{fig:gr.eps}
\end{figure}

Figure \ref{fig:gr.eps} compares exact values of $\psi(\langle a\rangle)$ and $g_R(\ag)$ given by Eqs.(\ref{eq:pmfinalII}) and (\ref{eq:odepmesh}) with approximate values obtained from Eqs.(\ref{eq:approxpgeo}) and (\ref{eq:approxgrgeo}) respectively, for three factors $\rho=\{0.5,1,2\}$. We note that, for $\Delta a / \bar{a} \lesssim 10^{-3}$, relative errors increase. This is due to the ``low dynamic'' of the variable $m$ as it approaches unity (and $\Delta a \rightarrow 0$). In this case, it is advisable to use $u$ and $v$ as the moduli of elliptic integrals instead of $m$ to restore the accuracy in Eq. \ref{eq:pmfinalII}.

The potential always has its minimum inside the cell at a certain radius $\rmin$ reagardless of the opening angle $\Delta \phi$. This radius is found by equating the right-hand side of Eq. \ref{eq:odepmesh} to zero. It depends on the parameters $\bara$, $\Delta a$, and $\rho$. We expect $\rmin < \bar{a}$ for large opening angles (or low azimuthal resolutions) and $\rmin > \bara$ for small ones (high azimuthal resolutions). When $\Delta \phi \rightarrow 0$, $\rmin \rightarrow \bara$ as the polar cell becomes a rectangle. Since $\bara \ne \ag$, there is in general a self-force at $R=\ag$ at high azimuthal resolution. However, the gravitational acceleration exceptionally cancels out if the shape factor has the remarkable value $\rhom \approx 3.530937746935705$. This is the single root of Eq. \ref{eq:approxgrgeo}. For $\rho = \rhom$, we have $\psi(\ag) \approx - 5.922555527732666 \, G \Sigma_0 \Delta a$. Thus, if the computational mesh $(\Delta a, \Delta \phi)$ contains cells all with shape factor $\rho = \rhom$, then there is no need to compute the gravitational force at $R=\ag$, since it is zero in the high resolution limit.

More generally, since the two moduli $m_1$ and $m_2$ appearing in Eq. \ref{eq:odepmesh} are functions of the ratio $\aout/\ain$, it is always possible to find a shape factor for which the acceleration vanishes at a fixed value of $R/\ain$. It can therefore be advantageous to build the polar mesh accordingly. For instance, let $N_R$ and $N_\phi$ be the numbers of cells in the radial and azimuthal directions, respectively, and $\rint \equiv R_1$ be the radius of the disc inner edge. Once the shape factor $\rho$ has been numerically determined such that $g_R=0$ (this is $\rhom$ in the high resolution limit), the radius $\rext = R_{{N_a}+1}$ of the outer edge is found from the relation:
\begin{equation}
\rext = \rint \times \left( \frac{N_\phi \rho + \pi}{N_\phi \rho - \pi} \right)^{N_R} \approx \rint \times \left(1 + \frac{2 \pi}{\rho} \frac{N_R}{N_\phi} \right)
\end{equation}
With such a mesh, there is no gravitational acceleration at the center of computational cells.

\section{Concluding remarks}

We have reported exact expressions for the self-gravitating potential and acceleration in homogeneous polar cells whatever their shape. The high resolution limit has been analyzed and reliable approximations have been derived. We confirm the validity of \cite{binneytremaine87}'s prescription for the singular term $K(0,0)$ to first order. We have shown that there is always a self-acceleration at the center of computational cells. However, this acceleration can cancel out at any radius by an appropriate choice of cell shape. This work\footnote{A Fortran 90 package called {\tt PolarCELL} is available at the following address:\\ {\tt www.obs.u-bordeaux1.fr/radio/JMHure/intro2polarcell.html}} contributes to the treatment of self-gravity in hydrodynamical simulations, and especially for discs.

\begin{acknowledgements}
It is a pleasure to thank  P. Barge, C. Baruteau and C. Surville for valuable discussions, as well as A. Jeffrey, co-editor of ``Tables of integrals, series and products'' by \cite{gradryz65}.
\end{acknowledgements}



\end{document}